# Site-Specific Contributions to the Band Inversion in a Topological Crystalline Insulator

D. Koumoulis[1], T.C. Chasapis[2], B. Leung[1], R.E. Taylor[1], C.C. Stoumpos[2], N.P. Calta[2], M.G. Kanatzidis[2,3], L.-S. Bouchard[1,4]

[1]*Dept. of Chemistry and Biochemistry, University of California, Los Angeles, CA 90095-1569 USA*

[2]*Dept. of Chemistry, Northwestern University, Evanston, Illinois 60208, USA*

[3]*Division of Materials Sciences, Argonne National Laboratory US DOE, Argonne, Illinois 60439, USA*

[4]*California NanoSystems Institute, UCLA, 570 Westwood Plaza, Los Angeles, CA 90095-1569 USA*

**In a topological crystalline insulator (TCI) the inversion of the bulk valence and conduction bands is a necessary condition to observe surface metallic states. SnTe is a well-known TCI with inverted band structure. Solid solutions of $Pb_{1-x}Sn_xTe$ have also been verified to be TCI, where band inversion occurs as a result of the band gap evolution upon alloying with Sn. The origins of this band inversion remain unclear. Herein we investigate the role of Sn insertion into the PbTe matrix for the *p*-type $Pb_{1-x}Sn_xTe$ series with *x*=0, 0.35, 0.60 and 1.00 via NMR and transport measurements. $^{207}Pb$, $^{119}Sn$ and $^{125}Te$ line shapes, spin-lattice relaxation rates and Knight shifts provided site-specific characterization of the electronic band structure. This probe of the electronic band structure shows that the band inversion is unaffected by lattice distortions but related to spatial electronic inhomogeneities formed by Sn incorporation into the PbTe matrix. Strong relativistic effects are found to be responsible for the band inversion, regardless of carrier type and concentration, suggesting a novel interpretation of the band gap evolution with composition. A semiconductor to metal-*like* transition is confirmed by the transition from Arrhenius to Korringa behavior over the entire temperature range near *x*=0.60, lying close to the band crossing region. The temperature dependences of the NMR parameters reveal a negative temperature coefficient of the direct gap for SnTe and positive coefficient for PbTe.**

# I. INTRODUCTION

Band inversion (BI) in a semiconductor was first reported by Dimmock, Melngailis, and Strauss [1] in 1966. They showed via band structure calculations and optical experiments that as the Sn fraction in the PbTe matrix increases, the band gap should continuously close at a fraction intermediate between PbTe and SnTe followed by a reopening of the gap. Since then, there have been several theoretical [2] and experimental investigations (electrical, thermal conductivity [3] and thermoelectric [4]) of the energy gap ($E_g$) dependence on the Sn fraction ($x$). Incidentally, the series $Pb_{1-x}Sn_xTe$ and $Pb_{1-x}Sn_xSe$ [5-6] are currently of considerable interest in condensed matter physics due to their topological crystalline insulator (TCI) property. TCI materials were predicted theoretically in 2011 [5] and studied experimentally via *A*ngle *R*esolved *P*hotoemission *S*pectroscopy (ARPES) [6] a year later. In $Pb_{1-x}Sn_xTe$ compounds, the crystalline symmetry plays a role in promoting the topological protection of the surface states. Bulk BI and strong spin-orbit coupling (SOC) are necessary conditions for the creation of gapless states at the metallic surfaces of a TCI. In PbTe, a direct band gap occurs at the *L* point in the Brillouin zone between a valence band (VB) maximum with $L_6^+$ symmetry and a conduction band (CB) minimum with $L_6^-$ symmetry. The same symmetries are shared with SnTe but the band symmetries are inverted relative to those of PbTe. First-principles calculations have shown that the $L_6^+$ band state is governed by *s* electrons contributed by the Pb and Sn atoms and *p* electrons contributed by the Te atoms. In contrast, Pb and Sn *p* electrons plus Te *s* electrons govern the $L_6^-$ symmetry. BI suggests the crossing of the aforementioned band edges at an intermediate tin composition in the PbTe matrix but the site-specific contribution of each element to the BI remain unknown.

The BI effect in TCIs has been studied experimentally by ARPES other optical techniques. However, both tehniques have significant limitations. ARPES studies of the BI effect rely indirectly on the observation of the Dirac cone evolution (a surface property) around the high symmetry point $\bar{X}$ and not directly on the BI process which is a bulk property. Also the band bending effect that results from surface affects the ARPES studies. A characteristic example is shown in Figure 1 that demonstrates the electronic structure of $Pb_{1-x}Sn_xTe$ (30 nm, 111) at the $\bar{\Gamma}$ point for $x$=0, 0.3, 0.75 and 1.00, as probed by ARPES. ARPES distinguished the formation of the Dirac cone, whose formation was observed at $x$=0.75, but failed to give indication for the development of the BI that occurs at lower Sn fractions. In addition, when it comes to TCIs, ARPES works best with *n*-type materials. ARPES studies of TCIs such as Pb$_{1-x}$Sn$_x$Te and SnTe, which are *p*-type materials, can only be done on doped versions to create *n*-type materials. The doping renders the observation of the BI ambiguous [12,13]. Optical absorption techniques were unable to quantify the contribution of Sn to the BI due to the Burstein-Moss shift observed in compositions with higher Sn content ($x \geq 0.25$) [1,6].

Nuclear magnetic resonance (NMR) techniques provide complementary information to the study of such materials. They can report on relativistic effects (such as SOC) related to heavy *Z*-elements in alloys [7-26]. Also, NMR can probe materials that are either *p*-type or *n*-type and also materials containing multiple fractions of different carrier types [12,13]. The influence of SOC on the $^{125}$Te and $^{77}$Se nuclear spin-lattice relaxation times ($T_1$) of the topological insulators (TIs) Bi$_2$Te$_3$ and Bi$_2$Se$_3$, respectively, was discussed in [26]. It is known that SOC is a necessary factor for promoting the BI process in TIs and TCIs. Recently, in a study of a topologically non-trivial half-Heusler compound (YPtBi) Nowak *et al.* [11] determined that a *negative* NMR resonance shift accompanied by a short $T_1$ is a universal key feature of BI in topologically nontrivial materials similar to Bi$_2$Te$_3$ [13, 26] and Bi$_2$Se$_3$ [26]. The NMR resonance shift is strongly dependent not only on the band structure features of each specific compound but also on both the carrier density and carrier type (*n*- or *p*-type). Unlike half-Heusler compounds (*e.g.,* YPtBi, YPdBi), the Pb$_{1-x}$Sn$_x$Te series do not contain unpaired 4*f* and *d*-electrons (and holes, as in Half-Heusler compounds) that affect both the magnitude and sign of the NMR shift due to paramagnetism.

In spite of the large number of investigations of lead chalcogenide systems thus far [27], there is still no direct experimental evidence of the site-specific contributions and the influence of the local disorder to the BI. To remedy this situation, we have carried out an NMR study of the BI mechanism in the Pb$_{1-x}$Sn$_x$Te system by separately probing the nuclear resonances from different atomic sites within the lattice. NMR is a local probe of structural and electronic properties in semiconducting alloys yielding microscopic and nanoscopic information by interrogating selected nuclei of interest [7]. By observing the resonances of the different nuclei, site-specific information about the contributions of local orbitals to the band structure can be extracted. While NMR has been used for the study of TIs [8-10], BI mechanisms in TCI materials have not yet been investigated.

Herein we study the effect of tin insertion on the non-trivial electronic properties of the *p*-type $Pb_{1-x}Sn_xTe$ TCIs by transport and NMR measurements. The room temperature Seebeck coefficient indicates a complex valence band structure and Fermi levels as deep as ~0.4 eV for hole concentrations in the range of $10^{20}$ cm$^{-3}$. Extensive NMR measurements were performed using $^{207}$Pb, $^{119}$Sn and $^{125}$Te as a probe of both the $L_6^-$ and $L_6^+$ bands of the Brillouin zone. By analyzing the static (lineshape) and dynamic (relaxation) NMR parameters we demonstrate that the *x*=0 and 0.35 compositions display a typical semiconducting behavior. On the other hand, the *x*=0.6 material, lying close to the band crossing region at room temperature, behaves as a nonmagnetic metal. The spin lattice relaxation, which gives an estimation of the Fermi level density of states, was found peaked at *x*=0.6, situation that might be attributed to strong relativistic effects and high effective masses due to band gap closing. The $1/T_1$ and resonance shift of $^{125}$Te in SnTe imply the case of a topologically non-trivial semiconductor with inverted band ordering relative to PbTe. Interestingly, a negative T-dependent energy gap at the *L* point of the Brillouin zone was revealed in the temperature range 120-400 K for *x*=1. Finally, the compositional and temperature dependencies of the NMR parameters suggest that the scenario of BI driven by structural disorder is incorrect. Based on our analysis, electronic inhomogeneity is the most probable driving force for the valence–conduction band inversion.

## II. EXPERIMENTAL DETAILS

*a) Synthesis*: Compounds from the series $Pb_{1-x}Sn_xTe$ were synthesized as polycrystalline ingots by direct combination of the elements. Appropriate ratios of Sn metal (*5N, American elements*), of Pb metal (*5N, American elements*), and Te metal (*5N, American elements*) were loaded in 13-mm OD fused silica ampoules. The ampoules were evacuated and flame sealed under a pressure of $10^{-4}$ mbar. The sealed ampoules were placed in a box furnace and heated to 1100 °C over a period of 12 h, dwelled at this temperature for 10 h and then slowly cooled to room temperature. Using the above method, pure polycrystalline ingots of the $Pb_{1-x}Sn_xTe$ were obtained as determined by powder diffraction.

*b) Crystal growth*: The growth of the $Pb_{1-x}Sn_xTe$ crystals was accomplished using the vertical Stockbarger-Bridgman method and it was identical for all the compositions reported here. A given composition of ~10 g was loaded in a tapered 10 mm OD fused silica ampoule and flame sealed under $10^{-4}$ mbar pressure. The hot zone was set to 1100 °C while the cold zone was set to 700 °C. The middle zone of the furnace was not powered up, thus allowing a natural temperature gradient between the hot and the cold zones. The temperature gradient generated by this configuration of the furnace was calculated to be 30 °C/cm in the crystallization temperature range of the compounds. The crystals were grown by passing the sealed ampoules through the furnace using a vertical translation motor, operating at a translation speed of 8.0 mm/h. A typical single crystal specimen obtained by this procedure is shown in Fig. 2a. The high crystalline quality of the final materials was confirmed by x-ray Laue diffraction patterns of shiny 001 surface cleaved along the cooling direction (Fig. 2b).

c) Powder *X*-ray diffraction patterns were collected on finely ground powders of the single crystals for all samples on a CPS120 Inel *X*-ray powder diffractometer using Ni-filtered Cu $K_\alpha$ radiation operating at 40 kV and 20 mA. A position sensitive detector was used in reflection geometry. The compositional variation of the lattice constants were obtained by single crystal diffraction data collected using a STOE IPDS2 diffractometer on pieces broken from the Stockbarger-Bridgman grown samples.

d) The Seebeck coefficient and electrical conductivity of ~ 2 mm × 3 mm × 8 mm parallelepiped samples were measured simultaneously under a helium atmosphere (~0.1 atm) from room temperature to 800 K using an ULVAC-RIKO ZEM-3 system. Room temperature Hall coefficients were measured with a home-built system in a magnetic field ranging from 0.5 to 1.25 T, utilizing simple four-contact Hall-bar geometry, in both negative and positive polarity of the magnetic field to eliminate Joule resistive errors.

e) NMR spectra and spin-lattice relaxation time ($T_1$) data were acquired using a Bruker DSX-300 spectrometer operating at frequencies of 62.79 MHz ($^{207}$Pb), 111.9 MHz ($^{119}$Sn) and 94.69 MHz ($^{125}$Te). The particle size was small enough to avoid RF skin-depth effects at these frequencies. A standard Bruker X-nucleus wideline probe with 5 mm solenoid coil was used. The $^{207}$Pb *π/2* pulse width was 4.5 µs, the $^{119}$Sn *π/2* pulse width was 2.5 µs and the $^{125}$Te *π/2* pulse width was 4 µs. The spectra were acquired using a spin-echo sequence [*π/2*)$_x$ – *τ* – *π*)$_y$- *τ* - *acquire*] with the echo delay, $\tau$, set to 20 μs or shortened to 10 μs for some of the $^{119}$Sn measurements due to the very short $T_1$ values. $T_1$ data were acquired with a saturation-recovery technique. A relaxation delay of at least five times $T_1$ was used in each experiment to allow full recovery of the magnetization. The $^{207}$Pb, $^{119}$Sn, and $^{125}$Te chemical shift scales were calibrated using the unified $\varXi$ scale [15], relating the nuclear resonance frequency to the $^1$H resonance of dilute tetramethylsilane in CDCl$_3$ at a frequency of 300.13 MHz. The reference compounds for assigning zero ppm to the various chemical shift scales are tetramethyllead for $^{207}$Pb, dimethyltelluride for $^{125}$Te, and tetramethyltin for $^{119}$Sn.

## III. RESULTS AND DISCUSSION

### A. Structural and transport properties

Figure 2c displays the powder *X*-ray diffraction patterns of all Pb$_{1-x}$Sn$_x$Te compositions where the 200 peak in the region ~ 27º is most intense and a systematic decrease in *d*-spacing with increasing Sn content is observed. The single crystal refinement confirmed that the compositions crystalize with the rock salt structure in the *Fm*-3*m* space group. The reduction of the lattice constant on increasing Sn fraction (Table I) is in accordance with the expected Vegard's law dependence for a solid solution (Fig. 2d). To rule out possible structural transitions, single crystal diffraction data of the Pb$_{0.65}$Sn$_{0.35}$Te and SnTe materials were collected in the range T=100-300 K (data not shown). In both cases, linear lattice shrinkage on cooling was observed, as expected.

The temperature-dependent Seebeck coefficients and electrical conductivities are shown in Figures 3a and 3b. The room temperature thermopower values are positive, indicating *p*-type conduction and they decrease with increasing Sn content from ~316 μV/K for PbTe to ~37 μV/K for SnTe. Room temperature electrical conductivities also scale with the Sn fraction from ~223 to ~7500 S/cm when going from *x*=0 to *x*=1. Both trends reflect the increase in hole density due to Sn vacancies, resulting in a deviation of the stoichiometric compositions towards the Te-rich side [28]. The room temperature hole densities of the Pb$_{1-x}$Sn$_x$Te compositions calculated by the respective Hall coefficients (Table I) show an increase of the free carrier density by two orders of magnitude when going from *x*=0 to *x*=1.

As illustrated in Fig. 3a, when raising the temperature, the Seebeck coefficients of the $x=0$ and $x=0.35$ compositions undergo a *p*- to *n*- transition. In particular, the thermopower increases when heating the sample up to ~ 450-500 K and decreases for higher temperatures, ending negative at 800 K. Furthermore, the respective electrical conductivities (Fig. 3b) decrease, accompanied by an upturn at temperatures of ~500-600 K. This behavior is attributed to the thermal excitation of the minority carriers, i.e., to the change of the conduction carriers from the extrinsic to the intrinsic state because of the low doping [29]. The thermally activated gap may be estimated from the Seebeck coefficient maxima through $E_g = \frac{S_{max}}{2 \cdot e \cdot T_{max}}$, where *e* is the elementary charge and $T_{max}$ is the temperature at which the maximum Seebeck occurs [30]. In this way, band gap values of ~0.37 eV and ~0.19 eV were obtained for the $x=0$ and $x=0.35$ respectively. These values lie close to those values determined by optical methods, supporting the expected closing of the band gap as a result of alloying PbTe with Sn.

For $x=0.60$ and $x=1$ compositions the hole densities are ~$10^{20}$ cm$^{-3}$. These materials are degenerate semiconductors without any bipolar diffusion. This is clearly demonstrated by the temperature dependence of the Seebeck coefficients of Fig. 3a, where increased thermopower values with temperature are observed without any sign of saturation or turnover. At the same time, the temperature dependence of the electrical conductivity denotes a metallic behavior because of high doping (see Fig. 3b).

Important information about the band structure of the Pb$_{1-x}$Sn$_x$Te series is obtained from the plot of the room temperature Seebeck coefficient against hole density, i.e., the Pisarenko plot [31] (Fig. 3d). As can be seen from Fig. 3d, the concentration dependent thermopower shows an anomalous behavior. It decreases from ~316 µV/K for the $x=0$ composition to ~20 µV/K for $x=0.60$ and then increases to ~37 µV/K for $x=1$. At the same time the hole density increases from ~$8 \times 10^{17}$ to ~$5 \times 10^{20}$ cm$^{-3}$ upon going from PbTe to SnTe (see Table I).

The solid lines of Fig. 3c represent the theoretical concentration dependence of the Seebeck coefficient on the assumption of a single parabolic band model, i.e., a single valence band located at the *L* point of the Brillouin zone that accounts for acoustic phonon scattering [32]. The theoretical lines for the $x=0$, 0.35, 0.60 and 1.00 compositions were obtained using for the density of states the effective masses $m^* \sim 0.4 m_0$, $0.21 m_0$, $0.17 m_0$ and $0.16 m_0$, respectively ($m_0$, mass of free electron; $m^*$, effective mass), as proposed by Efimova et al. [33] after the analysis of lightly doped $Pb_{1-x}Sn_xTe$ compositions with the same $x$ values. As can be seen from Fig. 3c, for the $x=0$ and $x=0.35$ materials a fairly good agreement between experiment and theory is obtained, meaning that a single band seems to dominate the room temperature transport properties. However, for the $x=0.60$ and $x=1$ compositions the experimental Seebeck values lie above the theoretical ones. In other words, for the two heavily doped compositions the transport behavior deviates from that predicted by a single band model.

The observed deviation may be explained by a complex valence band structure where, in addition the primary $L$ point maximum, a secondary lower lying one is located at the $\Sigma$ point of the Brillouin zone [34]. This highly degenerate maximum, the so-called heavy hole band, dominates the transport properties of strongly doped $p$-type PbTe [35], SnTe [36-38] and their solid solutions [33, 39]. For Fermi levels within a few $k_BT$ from the secondary maximum, increased Seebeck values are observed due to this high mass second valence band. In the case of a two band model the total thermopower is determined by the energy gap or the valence band offset between the two inequivalent maxima and the ratio of the effective masses and the mobilities for the two valence bands. In order to demonstrate the effect of the complex valence band structure on the transport properties, the concentration dependent thermopower of SnTe within the framework of a two band model [38] is presented in Fig. 3c. For SnTe the valence band offset is ~0.35-0.4 eV at 300 K and the contribution of the heavy hole band having $m^*\sim1.9m_0$ is located at hole densities $> 2\times10^{20}$ cm$^{-3}$. Due to the contribution of the heavy hole band, the Seebeck coefficient undergoes an increase, reaching a maximum for $\sim5\times10^{20}$ cm$^{-3}$ holes [38]. For PbTe the room temperature valence band offset is lower (~0.12 eV) and the Seebeck coefficient saturates due to the secondary extremum for hole densities $\sim5\times10^{19}$ cm$^{-3}$ [35].

It is clear from Fig. 3c that for the $x=1$ composition the experimentally measured Seebeck coefficient lies close to the one predicted by the two band model. For the $x=0.60$ material the application of the two band model is challenging because the dependence of the valence band offset on composition $x$ is not known at ambient temperatures [40,41]. Based on the band inversion model, the $x=0.60$ composition lies close to the band-crossing region [3,1]. This situation is expected to result in significant band non-parabolicity effects that may lead to an anomalous carrier concentration dependence of the density of states on effective mass as demonstrated in the case of *n*-type $Pb_{1-x}Sn_x$Te with $x > 0.5$ [39].

Based on the concentration dependent thermopower, we estimated the Fermi level as a function of hole density for the four studied compositions. For the $x=0$ and $x=0.35$ the Fermi level was found from the comparison of the experimental Seebeck values with the theoretical ones as predicted by the single band model (solid lines of Fig. 3c). Similarly, for the $x=1$ the Fermi level was obtained on the assumption of a two band model (dashed line of Fig. 3c). For the $x=0.60$ composition the respective value was determined by assuming a smooth variation of the Fermi level with hole density. The results are displayed in Fig. 3d. For the $x=0$ material with hole density of $\sim 10^{17}$ cm$^{-3}$ the Fermi level is located within the forbidden band width. For the $x=0.35$ composition having a hole density of $10^{18}$ cm$^{-3}$ holes the Fermi level is within the valence band, a few $k_B$T lower than the band edge. However, for the $x = 0.60$ and $x = 0.1$ samples with hole densities $\sim 10^{20}$ cm$^{-3}$, the Fermi level is located $\sim 0.3$ and $\sim 0.4$ eV deeper than the valence band edge, respectively, causing contributions from the heavy hole band at room temperature.

## B. NMR spectral analysis

The sign and magnitude of the NMR shift in semiconductors are related to the band structure, the type of semiconductor (*n*- or *p*- type) and the carrier density of the compound. The observed resonance shift is the sum of the chemical and Knight shift. The former is related to the interaction of the nucleus with filled electronic states including bonding electrons. Its magnitude is related to the changes induced by modifications of the bonding properties such as doping and local disorder. The latter is connected to the hyperfine coupling of the nuclear spin with the "free" charge carriers (delocalized electrons or holes). The net shift enables us to track the position and changes of the Fermi level as the temperature and Sn content in the Pb-Te matrix is varied.

### i. Doping dependence of the NMR spectra

The $Pb_{1-x}Sn_x$Te compounds contain three NMR-active nuclei: $^{207}$Pb, $^{125}$Te and $^{119}$Sn. The $^{207}$Pb lineshapes of $Pb_{1-x}Sn_x$Te obtained at 296 K are shown in Fig. 4a. The acquisition of each spectra required the application of several irradiation frequencies [14] due to their extended resonance shift range, which could not be uniformly excited with a single *RF* pulse. As the Sn content increases in the PbTe matrix, the $^{207}$Pb peak initially shifts upfield (lower frequency) from 1032 ppm for PbTe to -1744 ppm in the case of Pb$_{0.65}$Sn$_{0.35}$Te, in agreement with the expectation of an increase in the carrier density (see Table I) [24]. Such a resonance frequency shift is considered large for a semiconductor. Previous studies have shown that the resonance shifts of $^{207}$Pb can extend up to 12,000 ppm (relative to tetramethyllead [19, 42]). The $^{207}$Pb Knight shifts, for example, are approximately 1% for a carrier concentration of $10^{19}$ cm$^{-3}$ in *n*- and *p*-type PbTe semiconductors. The two components of the overall shift can be separated by extrapolating the resonance position to the limit of zero free carriers [12, 21]. If the carrier concentration (Knight shift) were the main contribution to the total shift, then we should expect the $^{207}$Pb spectrum of *x*=0.60 to be more upfield (lower frequency) than the spectrum of *x*=0.35. Since transport properties indicate that the hole density increases by two orders of magnitude on going from *x*=0.35 to *x*=0.60 (see Table I), the observed resonance shift must be primarily due differences in the chemical shift, not the Knight shift.

Another interesting finding is a double-peaked lineshape in the case of Pb$_{0.4}$Sn$_{0.6}$Te, where the $^{207}$Pb resonance frequency was near 18.2 ppm. The observed $^{207}$Pb Knight shift is almost an order of magnitude larger than those of $^{119}$Sn and $^{125}$Te as will be discussed later (*c.f.* Fig. 4b, c). We measured the frequency dependence of $T_1$ across the lineshape at ambient temperature. The Pb$_{0.4}$Sn$_{0.6}$Te sample has very short $^{207}$Pb $T_1$ values with $T_1$ decreasing with increasing frequency across the resonance. Specifically, the peak at 479 ppm has a $T_1$ equal to 321 μs whereas that of the second peak at -490 ppm is 480 μs. The different $T_1$ values across the lineshape accompanied by an asymmetric resonance are consistent with the presence of an inhomogeneous distribution of defects arising from different charge carrier concentrations located in different microcrystalline regions, resulting in a range of Knight shifts for nuclear spins located in these different environments throughout the sample. This well-known effect is often referred to as electronic inhomogeneity. Similar results have also been observed in previous NMR studies of PbTe [12,18] and PbSe [13].

Taking into account the fact that the incorporation of Sn in PbTe affects the lattice, we have investigated the dependence of $^{207}$Pb linewidth on Sn fraction. The $^{207}$Pb spectrum (Fig. 4a) broadens as the Sn content increases in the PbTe matrix. In addition, (at least) two individual lines appear in the $^{207}$Pb spectra. The $^{207}$Pb linewidth increases continuously from 340 ppm in Pb$_{0.65}$Sn$_{0.35}$Te to 650 ppm in Pb$_{0.4}$Sn$_{0.6}$Te. The presence of structural disorder in cubic materials is known to further broaden the NMR resonance of quadrupolar nuclei ($I > 1/2$) compared to the case of purely dipolar nuclei ($I = 1/2$). The absence of quadrupolar nuclei in the Pb$_{1-x}$Sn$_x$Te compounds, as well as the inability of magic angle spinning (MAS) to substantially narrow the PbTe [12] spectra (similar results have been observed for PbSe [13]), rules out both quadrupolar broadening (due to the presence of spin-*1/2* nuclei) as well as any broadening arising from anisotropic interactions (e.g., chemical or Knight shift anisotropy) [12, 13]. Therefore, the asymmetrical features observed in the lineshapes arise from an extended distribution of Knight shifts, indicating local inhomogeneity in carrier concentration. Dipolar or orbital terms may contribute to the linewidth since it is a material with *p*-type charge carriers [12]; however, these are not the dominant terms due to the high doping effect in the Pb$_{1-x}$Sn$_x$Te series.

The $^{119}$Sn spectra (Fig. 3b) show that as the Sn content increases, the lines shift toward positive frequencies. There is an abrupt shift from an upfield regime of -1600 ppm observed for Pb$_{0.65}$Sn$_{0.35}$Te to the downfield resonance near the 2000 ppm resonance observed for Pb$_{0.4}$Sn$_{0.6}$Te, which appears close to the resonance of SnTe. The NMR shift drastically increases with Sn doping while maintaining a symmetric shape over the entire doping range. However, it should be noted that due to short $T_1$ (μsec), the entire spectrum (see section C) couldn't be excited. The symmetric $^{119}$Sn lines indicate that there is no noticeable anisotropic shift at the Sn sites similar to that observed at the $^{207}$Pb or $^{125}$Te sites (see below). Furthermore, the difference in the $^{119}$Sn resonance shift between Pb$_{0.4}$Sn$_{0.6}$Te and SnTe is small compared to the difference between Pb$_{0.65}$Sn$_{0.35}$Te and Pb$_{0.4}$Sn$_{0.6}$Te but consistent with an increase in the hole density (see also Table I). The hole density is of the same order of magnitude for $x$=0.60 and $x$=1 but increases by two orders of magnitude on going from $x$=0.35 to $x$=0.60 (see Table I). Taken together, these results suggest that the electronic inhomogeneities that influence the $^{207}$Pb shifts do not affect the $^{119}$Sn spectra.

Turning our attention to the $^{125}$Te spectra, the $^{125}$Te resonances (Fig. 4c) show that as the Sn fraction increases in the PbTe matrix, the resonance shifts downfield from -1159 ppm to 655 ppm, in line with the room temperature transport results demonstrating a increase in carrier density (Fig. 3c and table I). Therefore, the magnitude of the $^{119}$Sn and $^{125}$Te NMR shifts could be dictated by the hyperfine coupling of the free charge carriers to the magnetic moment of the

$^{125}$Te and $^{119}$Sn nuclei. On the other hand, all samples are *p*-type semiconductors, according to the Seebeck coefficient analysis [43]. From the NMR standpoint, the resulting holes that couple to the $^{125}$Te nucleus through the hyperfine interaction, would be expected to produce a negative frequency shift [21, 24]. Instead, we observed a positive trend. Therefore, our observation that the $^{125}$Te and $^{119}$Sn shifts move in opposite directions as a function of Sn concentration is unexpected. A plausible reason could be that the magnitude of the chemical shift is larger than the Knight shift and dominates the total resonance shift since the band symmetries of SnTe are inverted in comparison to the PbTe [16]. By inspection in Fig. 4c for the $^{125}$Te resonance of PbTe (Fig. 4c), the line shape is relatively symmetric with a center frequency of -1,176 ppm. The resonance covers a range from roughly -1,150 ppm to -1,225 ppm. In this case, it is difficult to ascertain the exact contribution of the chemical shift anisotropy to the total linewidth; but a rough estimate is -1,150 ppm. This corresponds to a chemical shift of -1,147 ppm at 77 K. Recently reported $^{125}$Te MAS-NMR spectra [16] of Pb$_{1-x}$Sn$_x$Te for *x*=0.05 and 0.3 show a positive shift (on the ppm scale) similar to our observation. However, the spin lattice relaxation ($T_1$) of $^{125}$Te in this study was larger than 1 s, which suggests the dominance of the chemical shift in contrast to the very short $T_1$ values of semiconductors that result from Knight shift (free carriers) effects [12, 26]. To estimate the chemical shift in the SnTe, two different samples with different carrier concentrations, were synthesized. We extrapolated the $^{125}$Te frequency shift values to a zero carrier concentration and the chemical shift position was found to be -889 ppm. Our chemical shift value is in excellent agreement with previous NMR (and MAS-NMR) experimental and theoretical studies [16] (following adjustment for the different chemical shift reference standard used in their experiments). Ramsey [17] showed that the chemical shift tensor at a nuclear site is a sum of two contributions, named the orbital chemical shift and the spin-orbit contribution to the chemical shift. The second term, called "spin-orbit shift" is important when dealing with materials consisting of high-*Z* elements. Specifically, this term arises from the orbital motion of Bloch electrons, which affects the spin susceptibility of these chalcogenides, hence the shift magnitude. Relativistic band calculations [44, 46] in Pb$_{1-x}$Sn$_x$Te suggest that by increasing the Sn fraction above 35% the reduction of spin-orbit coupling is estimated to change from 0.50 (PbTe) to 0.23 (SnTe). This variation in the spin-orbit coupling contributes to the $^{125}$Te and $^{119}$Sn shifts governed by the spin-orbit shift and likely causes the

opposite behavior in the directions of the observed resonance shift in PbTe and SnTe as well as an opposite change in the value of $T_1$.

The structural and transport data have shown that the introduction of Sn into the matrix results in the occupation of Pb sites and the *p*-type doping is a result of Sn vacancies. This is in accordance with the changes in the $^{125}$Te linewidth, which undergoes significant broadening with increasing Sn fraction because of the increasing carrier concentration. In order to understand the correlation between the line broadening and conductivity we examined the case of two SnTe samples containing (see Fig. 4c for details) different carrier densities. As shown in Fig. 4c the magnitude of the resonance shift is different for the two samples, illustrating the impact of differences in the average electron density at the $^{125}$Te nucleus, $\langle |u_{\vec{k}}|^2 \rangle_{E_o}$ even in samples that have identical band-edge properties. Clearly as shown above, there is sample variability but the presence of native defects and impurities play the key role in the correlation between the line broadening and conductivity, thus providing a reasonable explanation of the sample dependence on carrier density [21].

### ii. Temperature dependence of the NMR spectra

Now we turn our focus to the T-dependences of the three NMR-active nuclei in Pb$_{1-x}$Sn$_x$Te. Figures 5a-c show the temperature dependence of the $^{207}$Pb NMR spectra of three samples above T=290 K. All $^{207}$Pb resonances shift downfield with increasing T, as expected for a *p*-type semiconductor [24]. Consequently, the variable temperature $^{207}$Pb spectra are in agreement with the Seebeck coefficients (Fig. 3c) that were examined over the same temperature regime. The $^{207}$Pb spectra for PbTe (6.2 ppm/K and 11.0 ppm/K) and Pb$_{0.65}$Sn$_{0.35}$Te (17.2 ppm/K) depend strongly on T denoting trivial semiconducting behavior (thermally activated carriers). For Pb$_{0.4}$Sn$_{0.6}$Te above 310 K, the spectra depend weakly on T, suggesting metallicity. Figure 6 shows the *T*-variation of the $^{207}$Pb resonance shift for Pb$_{0.65}$Sn$_{0.35}$Te across the entire T range. Upon cooling the resonance shift moves upfield. Similar behavior has been observed *via* $^{125}$Te NMR on the same compound (see below Fig. 8a,b) across the entire T range (160-423 K).

Figure 7 presents $^{119}$Sn resonances for Pb$_{0.65}$Sn$_{0.35}$Te. The $T_1$ was so short (e.g., 90 μs) that significant signal was lost during the echo. Consequently, studying the $^{119}$Sn NMR spectra at various temperatures required acquisition times of more than 24 hours. All spectra could be fitted to a single Lorentzian line (Fig. 7a) unlike the complex $^{207}$Pb spectra, which were not Lorentzian. $^{119}$Sn NMR resonances range from -1423 ppm at 296 K to -984 ppm at 360 K (see Fig. 7b). The T dependence of the $^{119}$Sn NMR spectra (Fig. 7b) matches that of the $^{207}$Pb spectra in Fig. 6 and confirms the *p*-type character (Fig. 3c) of this compound. However, for the Pb$_{0.65}$Sn$_{0.35}$Te compound the $^{119}$Sn spectral shift (6.90 ppm/K) as a function of T is almost three times smaller than for $^{207}$Pb. Its linewidth is 4.5 times greater than the $^{207}$Pb resonance at 296 K ($^{119}$Sn has a linewidth of ~2,700 ppm whereas $^{207}$Pb linewidth is ~611.55 ppm at 296 K). This result can be associated with an extensive allocation of local electronic states at the $^{119}$Sn nucleus environment due to the presence of Sn atoms not only in Pb sites but also to interstitial lattice positions [45].

Next, we studied the T-dependence of the $^{125}$Te NMR signals. The $^{125}$Te spectral parameters are shown in Figures 8a and 8b. At each temperature, the line shift depends on the Sn concentration (Fig. 8a). Specifically, with decreasing T, the resonance for $x \leq 0.35$ shifts upfield and a cusp appears at 250 K. The $^{125}$Te resonance for $x=0.60$ changes from T-independent to strongly T-dependent, shifting upfield upon increasing temperature (Fig. 8d) for SnTe ($x=1$). This is an interesting result given that the SnTe is also a *p*-type semiconductor (with a lower Seebeck coefficient than PbTe) and this sort of temperature dependence is usually observed for *n*-type semiconductors, not *p*-type [12, 13, 20-26]. This result is in line with the also unexpected *x*-dependence of the $^{125}$Te and $^{119}$Sn resonances that found on going from PbTe to SnTe. The sign of the resonance shift correlates not only with the type of carriers but also with the symmetry of their band edges. The band symmetries of SnTe are inverted relative to the PbTe and the temperature coefficient of the $^{125}$Te resonance shift (Fig. 8 a & d) directly unveils the transformation of the VB from $L_6^+$ symmetry for the PbTe (consistent with a positive temperature coefficient) to $L_6^-$ symmetry for the SnTe (negative temperature coefficient). The T-independent shift observed in the case of $x=0.60$ arises from the Fermi-contact term (Knight shift), related to the contact electrons (as confirmed by a Korringa behavior of the relaxation rates, see below). At this Sn fraction the CB and VB are now overlapped, yielding an increase in the DOS at the Fermi level and a metallic character due to the $E_g \rightarrow 0$ (Fig. 8a).

From the slopes of plots of the Knight shift ($K$) versus the susceptibility with T as an implicit variable (Fig. 8c), we can estimate the hyperfine coupling constant ($A_{hf}$) at each composition, by assuming that the contact term is the dominant one. The hyperfine field is defined by $H_{hf} = A_{hf} \cdot I \cdot \langle S \rangle$ where $I$ is the nuclear spin and $\langle S \rangle$ is the average electronic spin. $A_{hf}$ is analogous to the average electron density at the nucleus $\langle |u_{\vec{k}}(0)|^2 \rangle_{E_o}$. The $A_{hf}$ was found to anomalously vary with the Sn fraction in the matrix. Particularly, we observed that as the Sn fraction increases, the $A_{hf}$ of $^{125}$Te changes from 22.89 kOe/$\mu_B$ for PbTe to 83.55 kOe/$\mu_B$ for $x$=0.35 and then decreases again close to 20.88 kOe/$\mu_B$ for $x$=0.60 ($\mu_B$, Bohr magneton). The hyperfine field constants of $^{207}$Pb were found to vary similarly to those of the $^{125}$Te nucleus, from 1142.67 kOe/$\mu_B$ for PbTe to 9466.12 kOe/$\mu_B$ for $x$=0.35 and then decreases again close to 322.53 kOe/$\mu_B$ for $x$=0.60. The hyperfine field constant of $^{119}$Sn was 2422.17 kOe/$\mu_B$ for $x$=0.35. The enhancement of $A_{hf}$ is additional direct evidence of the electronic density modifications during the evolution of the BI that occurs around $x$=0.60 composition. The observed fourfold enhancement of $A_{hf}$ at $x$=0.35 reveals that electronic modifications in the band structure of Pb$_{1-x}$Sn$_x$Te take place earlier than the evolution of the band inversion. A previous study on Pb$_{1-x}$Sn$_x$Te [24] hastened to dismiss the study of the $^{125}$Te nucleus due to its weak hyperfine coupling. In contrast, our experimental results indicate that the $A_{hf}$ of $^{125}$Te is large and comparable to the heavier elements such as $^{207}$Pb. Additionally, the variation in the $A_{hf}$ of $^{125}$Te follows the $A_{hf}$ of $^{207}$Pb as Sn replaces Pb. The surprisingly large values of $A_{hf}$ for the $^{209}$Pb and $^{119}$Sn are in accordance with the large values of the orbital hyperfine couplings in the Pb$_{1-x}$Sn$_x$Te due to the large spin-orbit coupling that occurs around $x$=0.35 composition.

### C. Spin-lattice relaxation rates

The $^{125}$Te, $^{119}$Sn and $^{207}$Pb spin-lattice relaxation times interrogate the spin dynamics at the site of each nucleus. In semiconductors with delocalized charge carriers, relaxation is dominated by spin-flip scattering with the conduction charge carriers. Therefore, the relaxation behavior is correlated with the Knight shift. According to Selbach *et al.* [25] the spin-lattice relaxation rate in semiconductors is proportional to $(m^*)^{\frac{3}{2}} \cdot n_e \cdot (k_B \cdot T)^{\frac{1}{2}}$ (Maxwell-Boltzmann gas) where $m^*$ is the effective mass, $n_e$ is the carrier density and $k_B$ is the Boltzmann constant.

In the case of metals (Fermi-Dirac gas), the $1/T_1$ rate is proportional to $(m^*)^2 \cdot n_e^{\frac{2}{3}} \cdot (k_B \cdot T)$. Consequently, a variable temperature spin-lattice relaxation study ($T_1$) investigates the presence of effective mass variations. By combining the $^{125}$Te SLR ($1/T_1$) and carrier concentration, the $T_1$ can be a valuable tool for characterizing the carrier concentrations in multi-component tellurides including electronically inhomogeneous compounds [18].

The $^{125}$Te magnetization recovery in $Pb_{1-x}Sn_xTe$ ($x > 0$) series can be fitted with a single exponential function, $M(t) = M_0(1 - e^{-t/T_1})$. As shown in Fig. 9a the $^{125}$Te relaxation time decreases as a function of the Sn doping. This suggests an increase in the DOS, strongly related to the band gap variation as the Sn content of the matrix increases. The contribution of relativistic effects, such as the SOC (related to the effective mass, $m^*$) occurs with an enhancement in the spin-lattice relaxation rate ($1/T_1$) as we have shown *via* previous NMR studies on topologically non-trivial and band inverted materials such as $Bi_2Se_3$ and $Bi_2Te_3$ [8,26].

The plot of $1/(T_1.T)$ versus Sn content, $x$, for all samples (Fig. 9b) shows remarkable features as the Sn concentration increases in PbTe. The $1/(T_1.T)$ product probes the $\vec{q}$-integrated imaginary part of the dynamical spin susceptibility and is associated with the DOS at the Fermi level. Initially, the $1/(T_1.T)$ rapidly increases from 0.001 s$^{-1}$K$^{-1}$ for PbTe to 0.085 s$^{-1}$K$^{-1}$ at $Pb_{0.65}Sn_{0.35}Te$. Then it displays a broad maximum of approximately 0.251 s$^{-1}$K$^{-1}$ closer to that of $Pb_{0.65}Sn_{0.35}Te$. By using the T-dependence of the SLR for each compound, we have extracted the thermally activated energy values ($\Delta E$) for the entire $Pb_{1-x}Sn_xTe$ series (Fig. 9b inset). A combination of the $1/(T_1.T)$ and the extracted $\Delta E$ value (Fig. 9c) indicates that the energy gap values inconsistently vary with the Sn content in PbTe. Moreover, the higher value of $1/(T_1.T)$ for $Pb_{0.4}Sn_{0.6}Te$ confirms that its Fermi-level DOS is larger than the other stoichiometries, in agreement with the transport data.

The spin lattice relaxation rate also gives an estimation of the Fermi level DOS that is proportional to the effective mass and the carrier density as expressed by $(m^*)^{\frac{3}{2}} \cdot n_e \cdot (k_B \cdot T)^{\frac{1}{2}}$. The incorporation of Sn in the PbTe matrix leads to an increased carrier density (see Table I) resulting in higher values of the $1/(T_1.T)$ product on going from PbTe to SnTe (Fig. 9b). However, the higher $1/(T_1.T)$ product for the $x=0.60$ relative to $x=1$ may only be explained by the assumption of a higher effective mass for the $Pb_{0.4}Sn_{0.6}Te$ composition. As we have already discussed in the transport properties section, the effective mass was found ~$0.17m_0$ for $x=0.60$ and ~$0.16m_0$ for $x=1$ on the assumption of a single parabolic band model [33]. On the other hand, the room temperature Pisarenko plot of Fig. 3c demonstrated the contribution of the heavy hole band for both the $x=0.60$ and $x=1$ materials. The origin of a higher effective mass for the $x=0.60$ composition relative to $x=1$ material may be explained by two possible reasons: (1) either by a lower valence band offset resulting in a stronger contribution from the heavy hole band in the case of the $x=0.60$ composition, (2) or by the incorporation of relativistic effects such as enhanced SOC and band non-parabolicity effects leading to higher effective mass due to the overlap of conduction and valence bands, as a result of the band inversion. Considering that the valence band offset was proposed to increase with the Sn content [39], peaking around $x\sim0.7$ [40], one may assume that the band crossing effect is the more likely explanation for the higher $1/(T_1.T)$ product for the $x=0.60$ composition displayed in Fig. 9b.

The T-dependence of the $^{125}$Te spin-lattice relaxation mechanisms (Fig. 9c-f) unravels the BI mechanism, as we now explain. The band gap energy ($E_g$) that is affected by the strength of the spin-orbit interaction [46, 47] is expected to significantly affect the $T_1$ mechanism as $x$ increases [26]. The NMR unveils a Korringa mechanism when the $L_6^+$ and $L_6^-$ overlap each other ($x=0.6$). The $^{125}$Te temperature dependencies of $T_1$ for PbTe [12] and the $x=0.35$ compounds indicate the presence of a thermally activated relaxation process. The first relaxation mechanism occurs in the low-temperature regime (T<250 K) and the second one becomes effective at the high-T regime (T>250 K). The absence of a stepwise drop in $1/T_1$ with increasing T rules out the existence a multi-gap scenario driven by structural disorder in $Pb_{1-x}Sn_xTe$ as was initially proposed [2]. Therefore, both of these thermally activated relaxation mechanisms are due to carrier excitations across the band edge states. The reported data as shown in Fig. 9, in the low-T region give activation energy of about 2.72 kJ/mol (28.20 meV) for this process. At higher T's, the dominant mechanism is characterized by activation energy of 5.38 kJ/mol (55.76 meV). However, for $^{207}$Pb, PbTe reveals an activation energy of 13.8 kJ/mol, whereas the $x=0.35$ compound has an activation energy almost half (5.28 kJ/mol) that of the Sn-free sample (Fig. 9a-d). Above $x=0.35$, the $^{125}$Te SLR results do not display an Arrhenius behavior with T. The $x=0.60$ obeys a linear T dependence (Fig. 9d). A Korringa-like behavior [$1/(T_1.T) = const.$] holds within experimental error throughout the entire T range (160 - 400 K), as shown in Fig. 9d for the case of $Pb_{0.4}Sn_{0.6}Te$. The absence of a thermally activated relaxation mechanism indicates that the Fermi level is not positioned in a gap (or pseudogap). In the case of degenerate semiconductors, the Korringa process is given as:

$$\frac{1}{T_1} = \frac{256 \cdot \pi^3}{9 \cdot \hbar} \cdot \frac{\gamma_n^2}{\gamma_e^2} \cdot n^2 \cdot \langle |u_{\vec{k}}|^4 \rangle_{E_o} \cdot \chi^2 \cdot k_B \cdot T \quad (1)$$

where $\chi$ represents the spin susceptibility, $n$ the carrier density and $\gamma_n^2$, $\gamma_e^2$ are the nuclear and electron magnetogyric ratio respectively. Therefore, the Korringa product is of the form: $\frac{1}{T_1 \cdot T \cdot K^2} = \frac{4 \cdot \pi \cdot k_B \cdot \gamma_n^2}{\hbar \cdot \gamma_e^2} \equiv \frac{1}{S_0}$. The theoretical value that we get for the $Pb_{0.4}Sn_{0.6}Te$ is equal to $(T_1 \cdot T \cdot K^2)_{theor.} = 2.60 \cdot 10^{-6}$ s.K. The value $(T_1 \cdot T \cdot K^2)_{exp.}$ as function of T is shown in Fig. 8d as inset. The Korringa product is constant, approximately $6.49 \cdot 10^{-7}$ s.K across the entire T range and smaller than the theoretical value. A constant value is expected in this case since the electron-electron correlations (many body effects) were not taken into account of the Korringa product. By using the resonance shift values and the $T_1 \cdot T$ product value of 2.95, we obtained $S_0 = 0.25$ ($S_0 \leq 1$, free electron state). The major contribution of non-$s$ electrons in the spin-lattice relaxation mechanism elucidates the $S_0 \leq 1$ [48]. We expand our analysis further by using the relaxation data of both $^{125}Te$ and $^{207}Pb$ at ambient temperature for the case of $Pb_{0.4}Sn_{0.6}Te$. From equation (1), we can estimate the electronic probability density at the $^{125}Te$, $^{119}Sn$ and $^{207}Pb$ nuclei. Particularly, we found that the ratio, $\frac{(T_1)_{Pb}}{(T_1)_{Te}} = \frac{\langle |u_{\vec{k}}(Te)|^4 \rangle_{E_o} \cdot \gamma_{Te}^2}{\langle |u_{\vec{k}}(Pb)|^4 \rangle_{E_o} \cdot \gamma_{Pb}^2}$ gives a value for the $\frac{\langle |u_{\vec{k}}(Te)|^4 \rangle_{E_o}}{\langle |u_{\vec{k}}(Pb)|^4 \rangle_{E_o}}$ product of $1.94 \cdot 10^{-2}$, since $\frac{(T_1)_{Pb}}{(T_1)_{Te}} = 0.045$ and $\frac{\gamma_{Te}^2}{\gamma_{Pb}^2} = 2.33$. Furthermore, the same product by using the $^{119}Sn$ $T_1$ data suggests that $\frac{\langle |u_{\vec{k}}(Te)|^4 \rangle_{E_o}}{\langle |u_{\vec{k}}(Sn)|^4 \rangle_{E_o}} = 10.04 \cdot 10^{-4}$, since now $\frac{(T_1)_{Sn}}{(T_1)_{Te}} = 7.52 \cdot 10^{-4}$ and $\frac{\gamma_{Te}^2}{\gamma_{Sn}^2} = 0.72$. As a consequence, the above results suggest that for $x_{Sn} = 0.6$, where the BI takes place, the Pb site has a carrier density that is approximately 52 times higher than at the Te site. The Sn site has a carrier density that is 996 times higher than at the Te site. The above values for the density of the electronic wavefunctions also reflect the strength of the hyperfine coupling $\langle |u_{\vec{k}}(0)|^2 \rangle_{E_o}$ for each nucleus.

The results indicate that the $^{125}$Te nucleus experiences a weaker hyperfine coupling as compared to the other two nuclei. Based on the systematic increase of hole density indicative of an enhanced metallic behavior due to the high doping, one would expect the Korringa process to be maintained even in SnTe with a higher value than in the case of $x$=0.60. Nevertheless, as shown in Figure 9e the Korringa mechanism does not apply in the case of SnTe. Thus, the T-dependence of the $^{125}$Te spin dynamics reveals mainly changes in the density of states that occur across the direct $L$-point band gap for each material. We should mention that the NMR results do not rule out the presence of a second valence band in SnTe, but we do not consider it as a crucial factor in the interpretation of the resonance shift and spin-lattice relaxation data.

The spin-lattice relaxation rate data suggest that an insulator-to-metal-like transition takes place in the Pb$_{1-x}$Sn$_x$Te series, as confirmed by the transformation from an Arrhenius behavior towards a Korringa law that holds across the entire T range for the $x$=0.60 composition. Apart from the change in the density of states, the T-dependent relaxivity data above the $x$=0.60 is inconsistent with a trivial semiconductor. Specifically, we found that SnTe has smaller activation energy at the high temperature regime relative to the low temperatures in contrast to the behavior observed in samples with tin fractions below 0.5. This observance in both the $1/T_1$ ($^{125}$Te) as well as the negative temperature coefficient of the $^{125}$Te resonance shift in SnTe, is direct evidence of an energy gap with a negative T-dependence at the $L$ of the Brillouin zone for SnTe compared to the positive T coefficient ($\frac{\partial E_g}{\partial T} > 0$) for PbTe [47]. This is also evidence that the band symmetry of SnTe is different from the PbTe. Particularly, as shown in Fig. 9e, $1/T_1$ of SnTe follows an expected semiconducting behavior characterized by two activation energies equal to 3.22 kJ/mol or 33.33 meV for T<300 K and 2.04 kJ/mol or 21.14 meV for T>300 K. That is, the valence band maximum of SnTe is now the $L_6^-$ since the band gap has been inverted relative to PbTe. As a result the relaxivity follows an opposite trend in PbTe, where the conduction band minimum has $L_6^-$ symmetry (Fig. 9f).

## IV. CONCLUSIONS

To summarize, we have provided evidence that the three NMR-active nuclei in Pb$_{1-x}$Sn$_x$Te serve as effective probes of different electronic and topological properties of its band structure. The $^{207}$Pb and $^{119}$Sn resonances are sensitive mainly to the $L_6^-$ band, which has $s$ character whereas $^{125}$Te NMR spectra reflect both the $L_6^-$ and $L_6^+$ bands of the Brillouin zone. The hyperfine coupling constant of both $^{125}$Te and $^{207}$Pb anomalously varied with the Sn fraction. The hyperfine coupling of $^{125}$Te is unexpectedly large and comparable to the higher $Z$ element ($^{207}$Pb) demonstrating the presence of a strong orbital hyperfine field in these compounds. The fourfold enhancement of hyperfine field coupling around $x$=0.35 reveals that electronic modifications in the band structure of Pb$_{1-x}$Sn$_x$Te take place earlier than the evolution of the band inversion. The $^{125}$Te NMR shifts and relaxation rates ($1/T_1$) for $x$=0 and 0.35 followed an Arrhenius behavior, typically of trivial semiconductors. The $^{125}$Te NMR shift for the $x$=0.60 compound is T-independent and the relaxation rates follows the Korringa law, two signatures which are indicative of a nonmagnetic metal. The $1/T_1$ and resonance shift of $^{125}$Te in SnTe imply the case of a topologically non-trivial semiconductor characterized by a negative T-dependent energy gap at the $L$ point of the Brillouin zone in contrast to the trivial case of PbTe. Apart from the $^{125}$Te NMR results, the T-dependent $^{207}$Pb data were also inconsistent with the Seebeck coefficient data, which further enhances the initial picture obtained via the $^{125}$Te results. All these results obtained from NMR demonstrate an inverted band ordering in SnTe relative to PbTe. Meanwhile, the absence of a multistep-like mechanism in $1/T_1$ with increasing temperature indicates that a multi-gap scenario driven by structural disorder is not valid for the Pb$_{1-x}$Sn$_x$Te; hence, the band inversion is unaffected by lattice distortions but related to electronic inhomogeneities, regardless of the carrier type and concentration.

## ACKNOWLEDGMENTS

This research was supported by the Defense Advanced Research Project Agency (DARPA), Award No. N66001-12-1-4034. We acknowledge the use of instruments at the Molecular Instrumentation Center (MIC) facility at UCLA.

## REFERENCES

[1] J. O. Dimmock et al., *Phys. Rev. Lett.* **16**, 1193 (1966)

[2] X. Gao and M.S. Daw, *Phys. Rev. B* **77**, 033103 (2008)

[3] J.R. Dixon and R.F. Bis, *Phys. Rev.* **176**, 942 (1968)

[4] M. Orihashi, Y. Noda, L.-D. Chen, T. Goto, T. Hirai, *J. Phys. Chem. Sol.* **61** 919-923 (2000)

[5] L. Fu, *Phys. Rev. Lett.* **106**, 106802 (2011)

[6] P. Dziawa *et al.*, *Nature Mat.* **11**, 3449 (2012)

[7] J.P. Yesinowski, *Top. Curr. Chem.* **306**, 229-312 (2012)

[8] D. Koumoulis *et al.*, *Phys. Rev. Lett.* **110**, 026602 (2013)

[9] B.L. Young et al., *Phys. Rev. B* **86**, 075137 (2012)

[10] Nisson et al., *Phys. Rev. B* **87**, 195202 (2013)

[11] B. Nowak and D. Kaczorowski, *J.Phys. Chem.C* **118** (31), 18021–18026 (2014)

[12] R.E. Taylor *et al.*, *J. Phys. Chem. C* **117**, 8959-8967 (2013)

[13] D. Koumoulis *et al.*, *Phys. Rev. B* **90**, 125201 (2014)

[14] D. Massiot, I. Farnan, N. Gautier, N. D.Trumeau, A. Trokiner, J.P. Coutures, *Solid State Nucl. Mag.* **4**, 241−248 (1995)

[15] R.K. Harris, E.D. Becker, S.M.C. De Menezes, R. Goodfellow, P. Granger, Pure Appl. Chem. **73**, 1795-1818 (2001)

[16] B. Njegic, E.M. Levin, K. Schmidt-Rohr, *Solid State Nuclear Magnetic Resonance* 55-56, 79-83 (2013)

[17] N.F. Ramsey, *Phys. Rev.* **77**, 567 (1950)

[18] E. M. Levin, J. P. Heremans, M. G. Kanatzidis, and K. Schmidt-Rohr, *Phys. Rev. B* **88**, 115211 (2013)

[19] C. Brevard, P. Granger. *Handbook of High Resolution Multinuclear NMR*, John Wiley and Sons, New York, 1981

[20] B. Sapoval, *J. Phys. (Paris) Suppl*. **29**, C4-133 - C4-136 (1976)

[21] S.D. Senturia, A.C. Smith, C.R. Hewes, J.A. Hofmann, P.S. Sagalyn, *Phys. Rev. B* **1**, 4045 – 4057 (1970)

[22] B.Sapoval, J.Y. Leloup, *Phys. Rev. B* **7**, 5272 – 5276 (1973)

[23] J.Y. Leloup, B. Sapoval, G. Martinez, *Phys. Rev. B* **7**, 5276 – 5284 (1973)

[24] C.R. Hewes, M.S. Adler, S.D. Senturia, *Phys. Rev. B* **7**, 5195 – 5211 (1973)

[25] H. Selbach *et al.*, *Phys. Rev. B* **19**, 4435 (1979)

[26] R.E. Taylor *et al.*, *J. Phys. Chem.* **116**, 17300-17305 (2012)

[27] Z. Zhu, Y. Cheng, and U. Schwingenschlögl, *Phys. Rev. B* **85**, 235401 (2012)

[28] R. F. Brebrick, *J. Phys. Chem. Solids*, 24, 27-36, (1963)

[29] M. Orihashia, Y. Nodab, L.-D. Chena, T. Gotoa, T. Hiraia, *Journal of Physics and Chemistry of Solids* 61 (2000) 919–923

[30] H.J. Goldsmid and J.W. Sharp, *Journal of Electronic Materials*, Vol. 28, 869 - 872, (1999)

[31] J.P. Heremans, B. Wiendlocha and A. M. Chamoire, Energy Environ. Sci., 5, 5510 (2012)

[32] A. F. May and G. J. Snyder, *Introduction to Modeling Thermoelectric Transport at High Temperatures, in Materials, Preparation, and Characterization in Thermoelectrics*, Edited by D. M. Rowe, CRC Press 2012, Pages 1–18

[33] B.A. Efimova and L.A. Kolomoets, *Sov. Phys. Solid State*, 7, 339 (1965)

[34] R.S. Allgaier and B. Houston, Phys. Rev. B, 5, 2186 (1972)

[35] S.V. Airapetyants, M.N. Vinogadova, I.N. Dubrovskaya, N.V. Kolomoets and I.M. Rudnik, *Sov. Phys. Solid State*, 8, 1069 (1966)

[36] R.F. Breberick and A.J. Strauss, *Phys. Rev.,* 131, 104, (1963)

[37] L. M. Rogers, *Brit. J. Appl. Phys. (J. Phys. D),* 1, 845 (1968)

[38] Q. Zhang, B. Liao, Y. Lan, K. Lukas, W. Liu, K. Esfarjani, C. Opeil, D. Broido, G. Chen, and Z. Ren, PNAS, 110, 13261–13266 (2013)

[39] M. Ocio, *Phys. Rev. B*, 10, 4274 (1974)

[40] A. Aladgadgyan and A. Toneva, *Phys. Stat. Solidi b*, 85, K131 (1978)

[41] E.A. Gurieva, I.N. Dubrivskaya and B.A. Efimova, *Sov. Phys. Semicond*, 4, 197 (1970)

[42] A. Bielecki, D.P. Burum, *J. Magn. Reson.* **116** (A) 215–220 (1995)

[43] F. Mott and H. Jones, *The theory of the properties of metals and alloys*, Clarendon Press, Oxford, (1936).

[44] F. Herman *et al. Journal de Physique Colloques* **29**, C4-62-C4-77 (1968)

[45] D. Shaw, *J. Phys. C: Solid State Phys.* **14** L869-L876 (1981)

[46] C. Niu *et al. Mater. Express* **3** 2 (2013)

[47] Y.W. Tsang and M.L. Cohen, *Phys. Rev. B* **3** (1971)

[48] M. Corti, A. Gabetta, M. Fanciulli, A. Svane, N.E. Christensen, *Phys. Rev.* **B** 67, 064416 (2003)

**Figures and Tables**

**Table I:**

**Room temperature transport and structural properties of the $Pb_{1-x}Sn_xTe$ compositions**

| Composition | Seebeck coefficient (μV/K) | Carrier concentration ($\times 10^{18}$ cm$^{-3}$) | Lattice Constant (Å) |
|---|---|---|---|
| PbTe | 317 | 0.8 | 6.45 |
| $Pb_{0.65}Sn_{0.35}Te$ | 106 | 7.5 | 6.40 |
| $Pb_{0.40}Sn_{0.60}Te$ | 17 | 145.6 | 6.36 |
| SnTe | 37 | 441.2 | 6.31 |

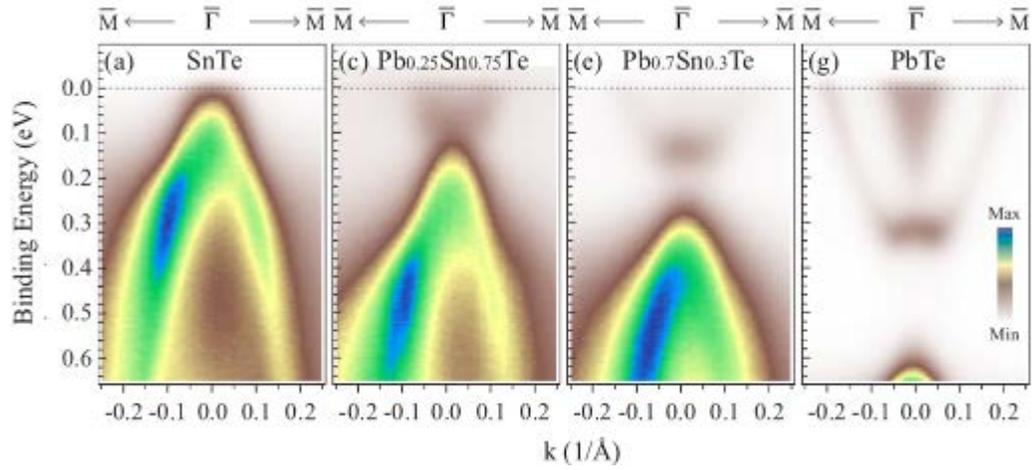

**Fig.1.** (Color online) Electronic structure of $Pb_{1-x}Sn_xTe$ (30 nm, 111) near the $\bar{\Gamma}$ point of $x=0$, 0.3, 0.75 and 1. The topological phase transition occurs between $x=0.3$-$0.75$ [figure reproduced with permission from *Phys. Rev. Lett.* **112**, 186801 (2014)].

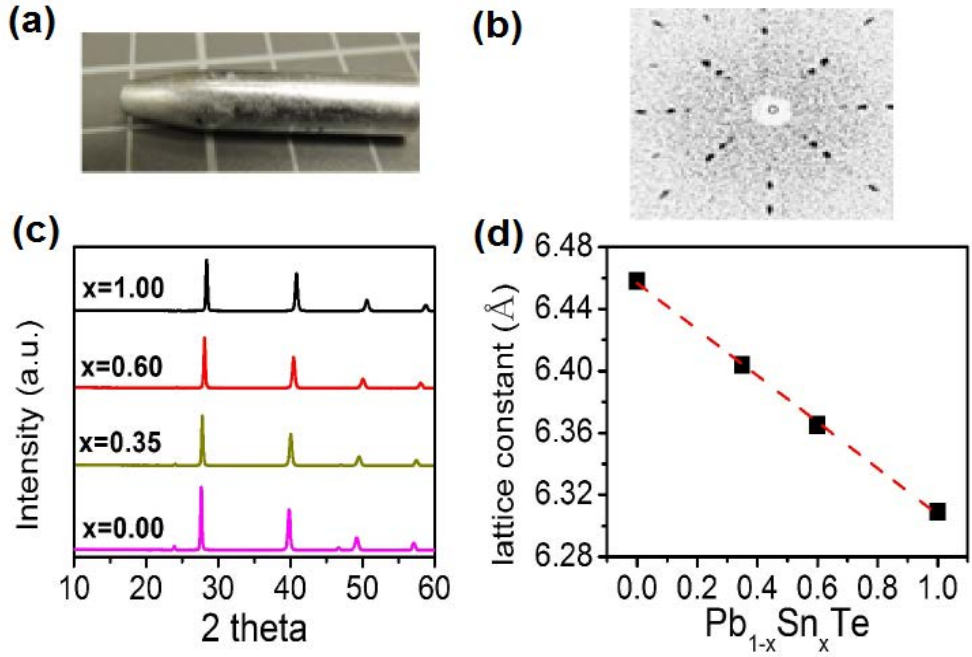

**Fig. 2.** (Color online) (a) Typical single crystal specimen obtained by using the vertical Stockbarger-Bridgman method, (b) Laue pattern of the final material reveal high crystalline quality samples, (c) Powder x-ray diffraction patterns of the studied compositions, (d) The compositional dependence of the lattice constant. Linear lattice shrinkage is observed upon alloying with SnTe.

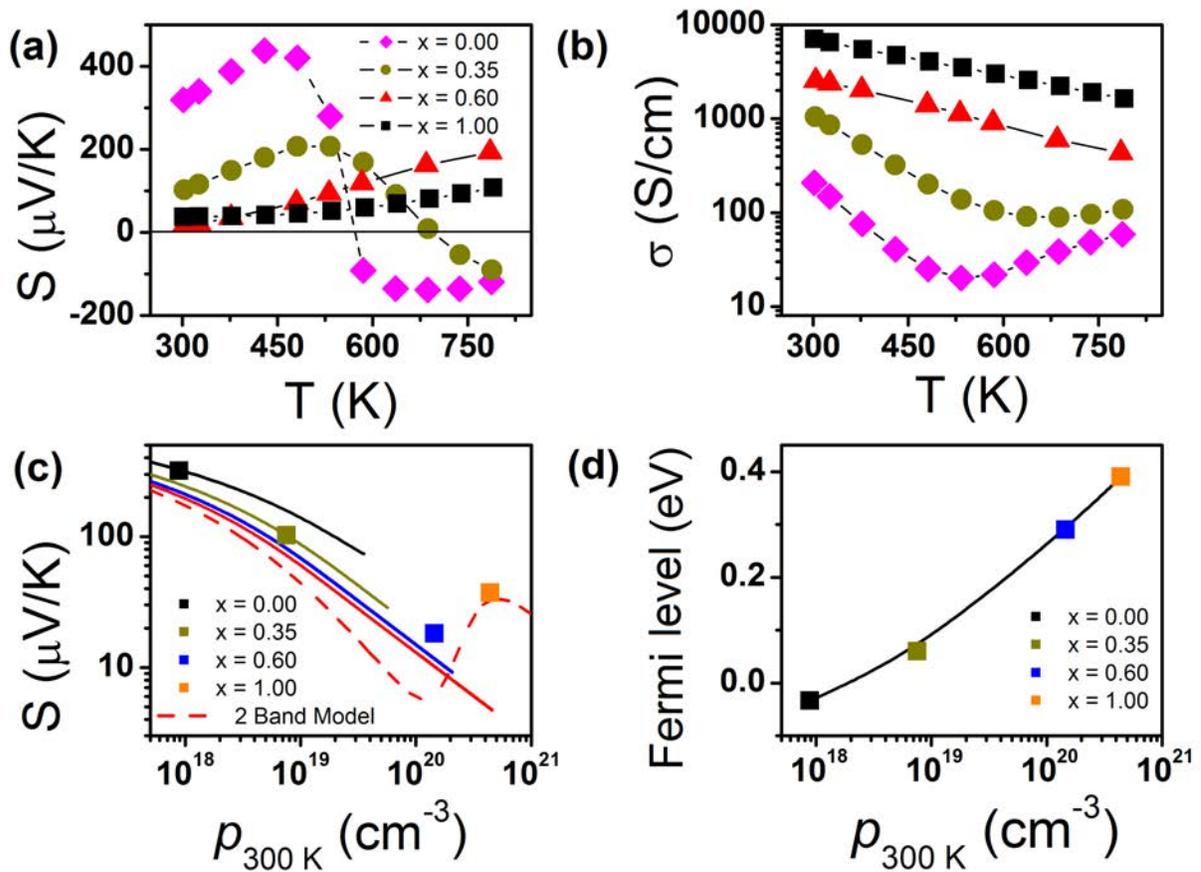

**Fig. 3.** (Color online) Temperature dependent Seebeck coefficient (a) and electrical conductivity (b) of the $Pb_{1-x}Sn_xTe$ with $x$=0, 0.35, 0.60 and 1.00. Note the onset of bipolar conduction at ~450 K and ~ 600 K for the $x$=0 and $x$=0.35 compositions respectively, (c) Room temperature Pisarenko plot. The solid lines are calculated on the assumption of a single parabolic band model. The dashed line represents the concentration dependent Seebeck coefficient of SnTe within the framework of a two-band model (see text for details), (d) Estimation of the Fermi level against hole density.

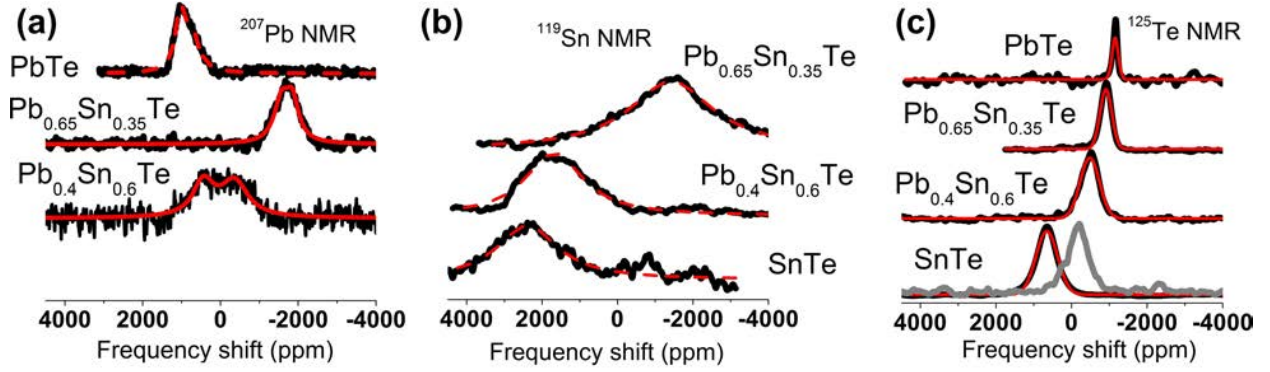

**Fig. 4.** (Color online) NMR frequency shifts of (a) $^{207}$Pb, (b) $^{119}$Sn and (c) $^{125}$Te measured at ambient temperature. One the one hand, the $^{207}$Pb line shape changes as the Sn doping increases with the shift varying initially negatively and then reversing again towards that of "pure" PbTe. The asymmetrical $^{207}$Pb spectra suggest the presence of regions with different carrier concentrations. Above $x=0.35$ the spectrum shifts to higher frequencies accompanied by an accession in the spread of the spectrum. The $^{207}$Pb spectrum of $x=0.60$ consists of two different peaks. On the other hand, with increasing Sn content, the observed positive shifts of $^{125}$Te and $^{119}$Sn spectra are accompanied by a broadening of the $^{125}$Te and $^{119}$Sn linewidths. The solid black line and grey line present the $^{125}$Te spectra of two different SnTe samples with different carrier concentrations. The solid black line refers to the $x=1$ material with $p_{300\ K} \sim 4.5 \times 10^{20}$ cm$^{-3}$ and the dotted line to a SnTe single crystal with lower hole density, $p_{300\ K} \sim 2 \times 10^{20}$ cm$^{-3}$. The sign of the $^{125}$Te shift is similar in both samples suggesting identical band ordering while the variation in the magnitude of the shifts illustrates the difference in the average electron density at the $^{125}$Te nucleus.

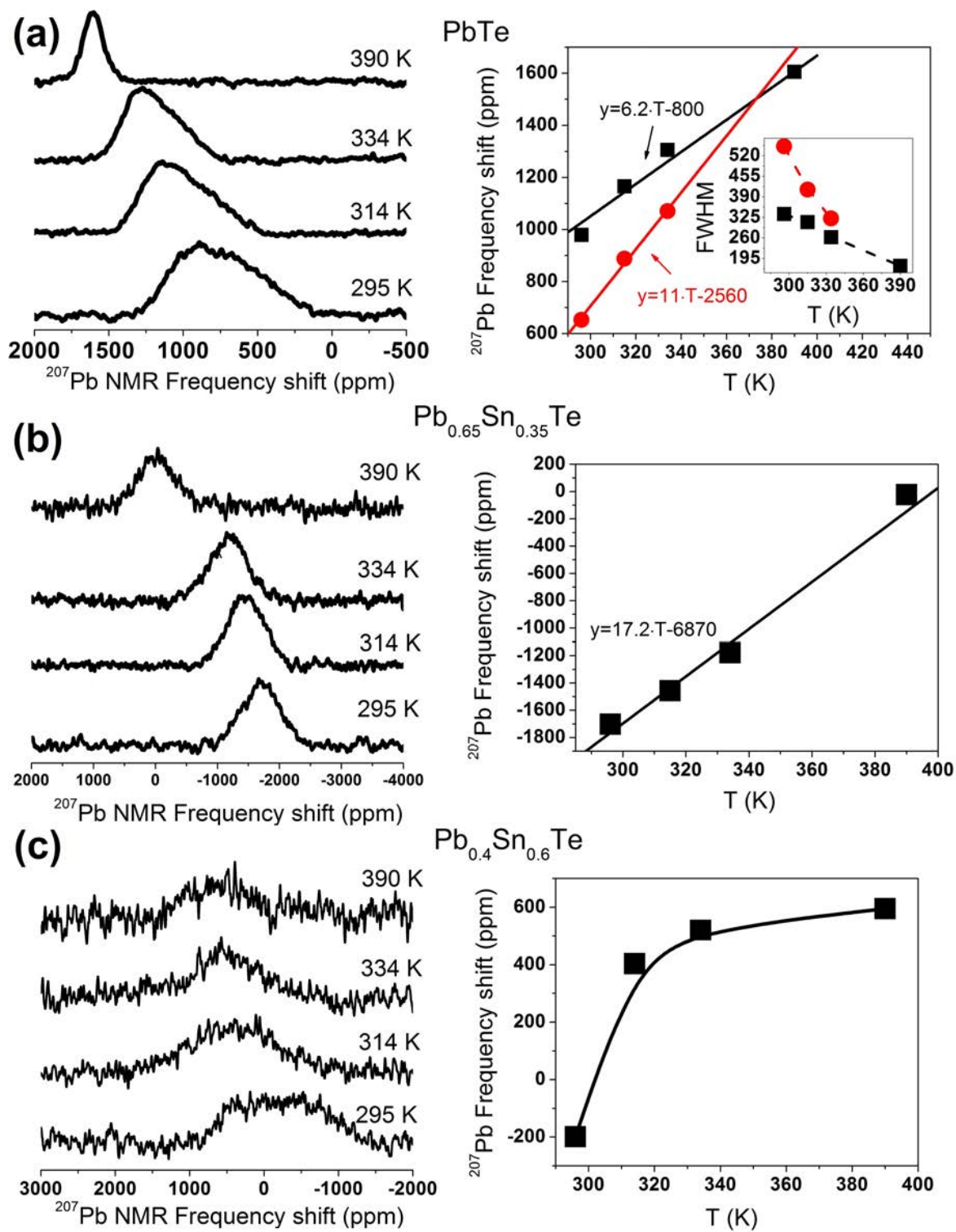

**Fig. 5.** (Color online) $^{207}$Pb line shapes for (a) PbTe, (b) Pb$_{0.65}$Sn$_{0.35}$Te and (c) Pb$_{0.4}$Sn$_{0.6}$Te at different temperatures.

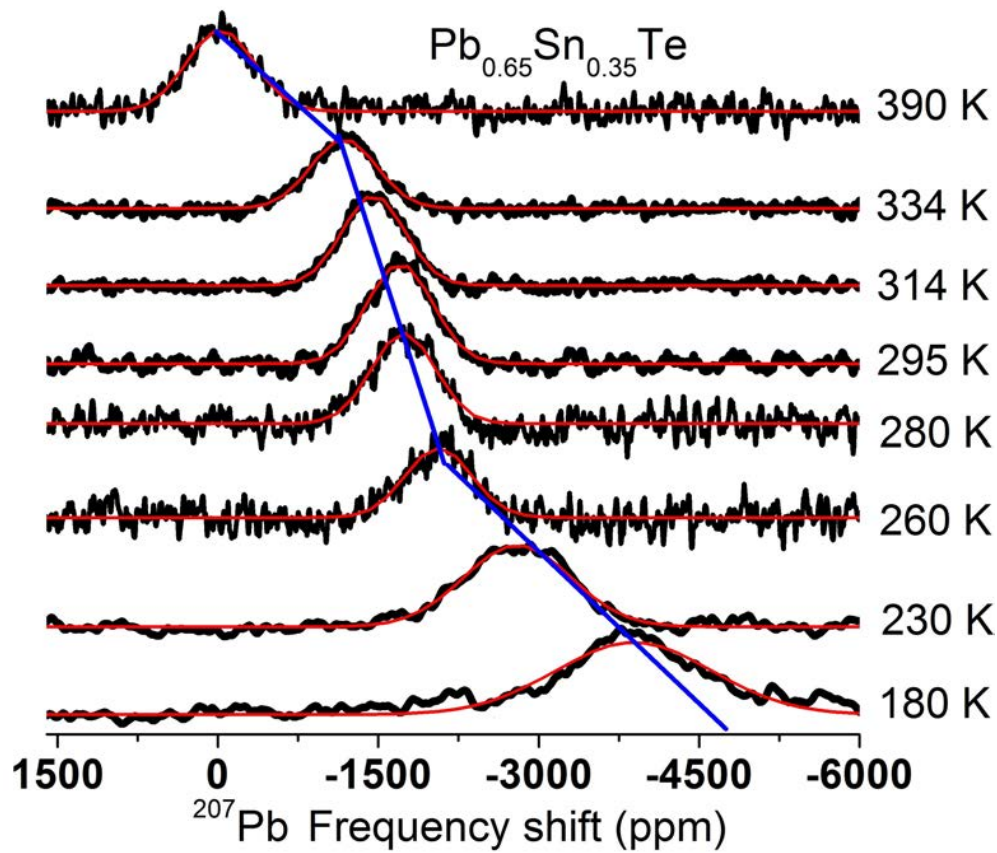

**Fig. 6.** (Color online) $^{207}$Pb spectra for Pb$_{0.65}$Sn$_{0.35}$Te across the temperature range 180-390 K. Below 250 K, the $^{207}$Pb linewidth increases and shifts upfield (approximately -4300 ppm).

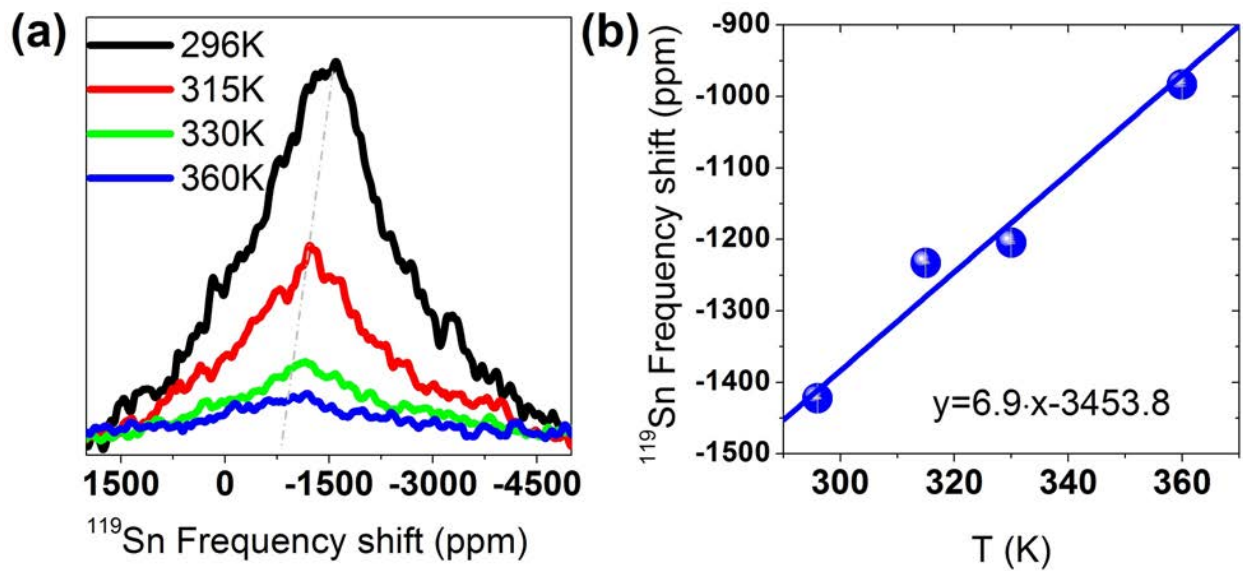

**Fig. 7.** (Color online) $^{119}$Sn spectra (a) and resonance shifts (b) as function of temperature for Pb$_{0.65}$Sn$_{0.35}$Te.

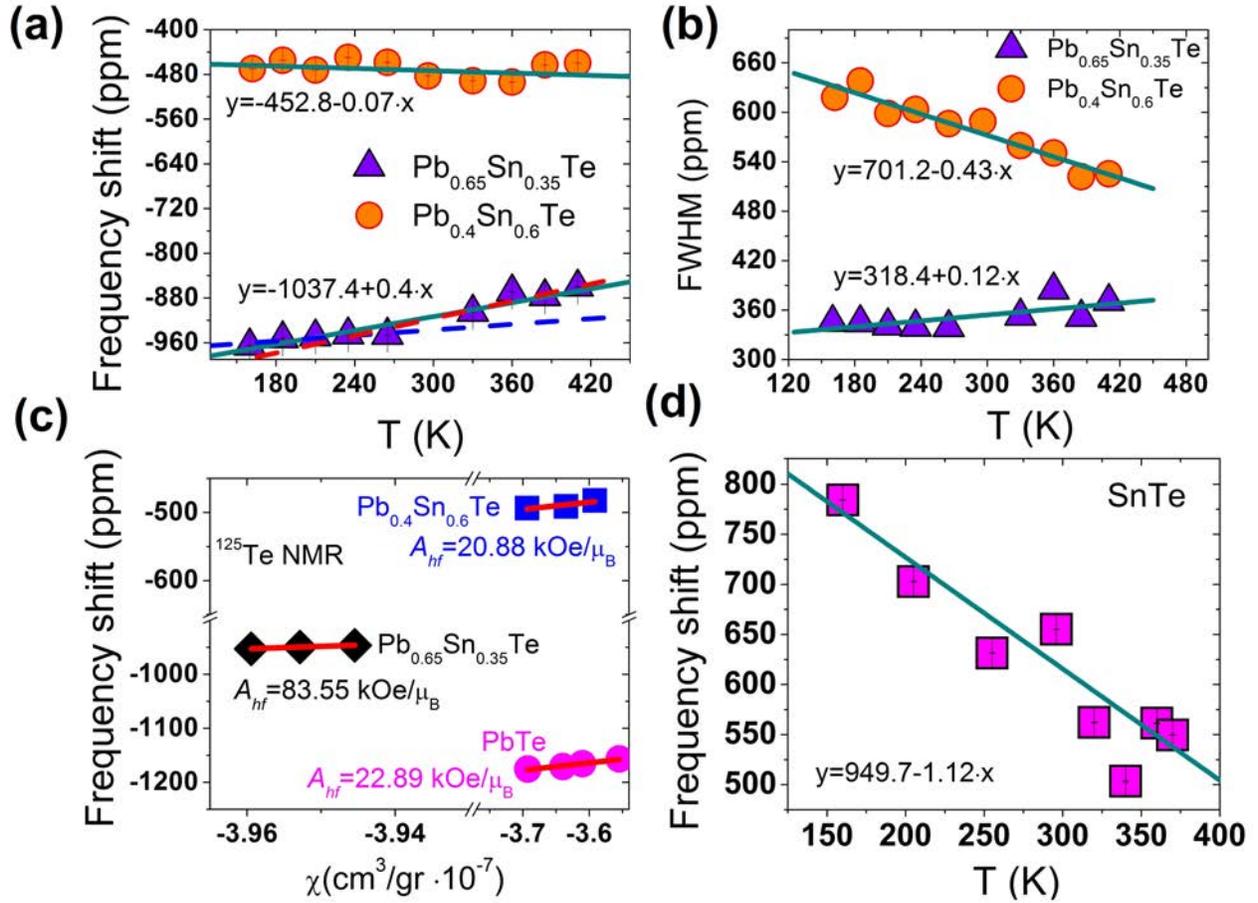

**Fig. 8.** (Color online) (a) Temperature dependence of $^{125}$Te resonance shift of $Pb_{0.65}Sn_{0.35}Te$ (purple triangles) and $Pb_{0.40}Sn_{0.60}Te$ (orange circles) from 160 K to 423 K (top). (b) T-dependence of $^{125}$Te linewidths of $Pb_{0.65}Sn_{0.35}Te$ (purple triangles) and $Pb_{0.40}Sn_{0.60}Te$ (orange circles) (c) Plots of the K-χ give the $^{125}$Te NMR hyperfine coupling constants of PbTe and the two alloys (x=0.35 and 0.60). (d) $^{125}$Te resonance shift in SnTe. As the temperature increases the spectrum shifts negatively from 780 ppm (160 K) to 540 ppm (370 K).

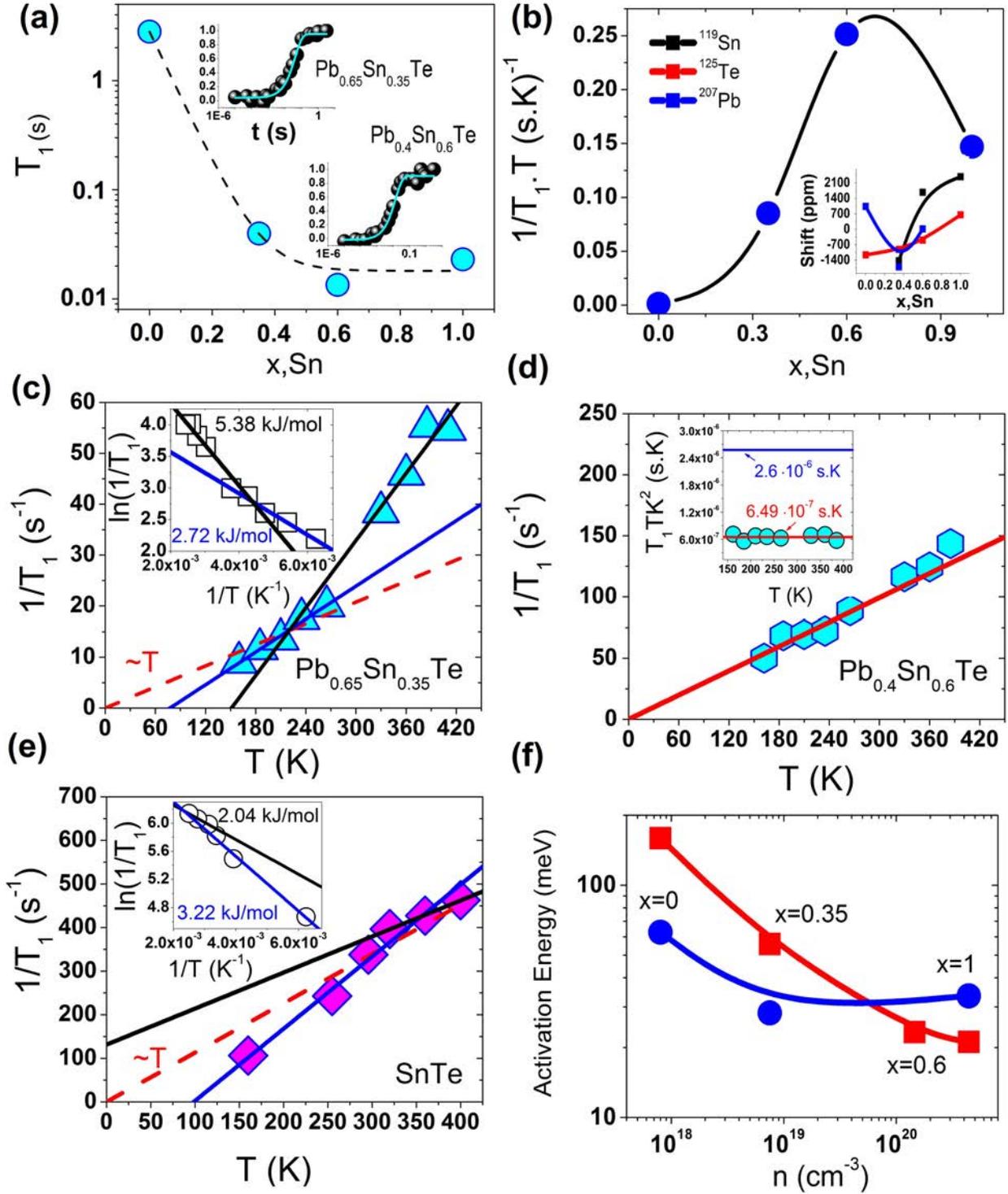

**Fig. 9.** (Color online) Temperature dependence of $^{125}$Te spin-lattice relaxation rate ($1/T_1$) in Pb$_{1-x}$Sn$_x$Te as function of doping range (a). The temperature dependence of the $1/(T_1 \cdot T)$ for $^{125}$Te probes the increase of the DOS at the Fermi surface as moved from PbTe to SnTe with a maximum peak appears at $x=0.60$ where the level-crossing occurs (b). The inset shows the doping dependence of the resonance shift for each nucleus. Temperature dependence of $^{125}$Te ($1/T_1$) of Pb$_{1-x}$Sn$_x$Te as function of temperature for $x=0.35$ and its natural logarithm of the $1/T_1$ (inset) as a function of the inverse temperature (c). The estimated thermal activation values ($E_g$) of the low temperature regime (blue line) and the high temperature regime (black line) are shown as insets. In the case of $x=0.60$ (d) the level crossing of the band edges is reflected in the $T_1$ study via the Korringa law (red line). Right after the inversion at the band edges, the system displays semiconducting behavior as it is reflected by the appearance of thermally activated process on the $1/T_1$ versus temperature of SnTe (e). The low (blue line) and high temperature (red line) thermal activation process for the Pb$_{1-x}$Sn$_x$Te series as function of Sn fraction in the PbTe matrix. The thermal activation energy decreases as the Sn fraction increases and reaches a minimum at $x=0.60$ (f).